# Heterophasic oscillations in nanometer-size systems

Alexander Patashinski and Mark Ratner

Northwestern University, Department of Chemistry

## Abstract

Singularities in macroscopic systems at discontinuous phase transitions are replaced in finite systems by sharp but continuous changes. Both the energy differences between metastable and stable phases and the energy barriers separating these phases decrease with decreasing particle number. Then, for small enough systems, random heterophasic oscillations of the entire system become an observable form of thermal motion. Under certain conditions, these oscillations take the form of oscillatory nucleation. We discuss mechanisms and observation conditions for these random transitions between phases.

In macroscopic systems, a discontinuous (first order) phase transition is manifested by discontinuity of thermodynamic properties at the transition temperature $T_\lambda$. For finite systems, these singularities are replaced [1-19] by a continuous change in a transition range $\Delta T_{tr}$ of temperatures. The transition range increases upon diminishing the number $N$ of particles, but it has been found rather narrow for systems with a large particle number: $\Delta T_{tr}/T_\lambda \ll 1$ for $N \gg 1$. This allows one to treat the rapid change as a smoothed phase transition, approximately characterized by a transition temperature. It has been known for a long time that this temperature shifts with diminishing system size [1-6]. Recent detailed studies of finite systems in the transition range of temperatures revealed a rather complex set of behaviors: surface melting, non-crystalline isomers, pattern formation, and properties fluctuations at time-scales much larger than particle vibration period [2, 3, 7–12].

Some of the observed phenomena may be explained by the increased role of the surface in smaller systems [3,7,8]. Surface-related effects can be controlled, for example by coating or placing the system in an appropriate matrix. Other new phenomena reflect features common to all finite systems: energy differences and excitation barriers between different thermodynamic states in finite systems are finite and decreasing when the particle number decreases. Then, heterophasic fluctuations that are local phenomena in macroscopic systems may change the state of the entire finite system. The objective of the current study is to discuss conditions for observing these fluctuations. We first consider a toroidal system that has no external surface; these systems are realized in computer simulations under periodic boundary conditions. Systems with an external boundary are then discussed on assumption that the system is large enough so that bulk prevails in determining the properties.

In recent decades, computer simulations became an increasingly important source of information about finite systems. Three-dimensional (3D) and two-dimensional (2D) systems of $N=(10^2-10^7)$ particles have been studied [2,3,9-20] using different simulation algorithms, frequently with the goal of understanding phase transitions in macroscopic systems. One of the advantages of the method is that simulation algorithms are designed to produce an equilibrium isochoric (NTV) or isobaric (NTP) Gibbs ensemble. For real systems, these standard experimental conditions are unambiguous only in the macroscopic limit; in smaller systems, the equilibrium ensemble may depend on the environment [18, 19].

One expects that at least in large enough finite systems kinetics of phase transformation is similar to the nucleation kinetics in macroscopic systems [20]: a change of state parameters from the old to new phase makes the old phase metastable, and then nuclei of new phase appear in the old phase as fluctuations, reach critical size, and grow to become macroscopic regions in the final equilibrium state. Transient states between the metastable old phase and the stable new phase are two-phase, with part of the system with $N_{new}<N$ particles in the new phase and the rest of the system with $N_{old} = N - N_{new}$ particles is in the old phase. The phases are separated by an interface; we consider here the case when the width of this interface is negligibly small compared to the system size. As a fluctuating mode, the number $N_{new}$ of particles in the new phase has a large relaxation time, so all other characteristics including the shape of the interface may be assumed equilibrated at given $N_{new}$. In this two-phase approach, the interface is characterized by surface tension $\alpha$ giving the energy cost of creating a unit area of the interface. In isotropic models, $\alpha$ is treated as a scalar function of state; when anisotropy is important, the surface tension become a function of a point at the interface depending on the local orientation of the interface relative to the anisotropy. Note that we label the phases "new" and "old" as a conventional way to distinguish between phases.

In an isobaric system at the phase transition point, the Gibbs free energies of the new and old phases coincide. In a macroscopic system, the energy barrier for a heterophasic fluctuation in the entire system is infinitely high, forbidding heterophasic fluctuations of the entire system at any time-scale. This barrier becomes finite for a finite supercooling or overheating, but then the energy difference between the old and the new phase in a macroscopic system is macroscopically large, making the transition from metastable to stable state irreversible. Heterophasic fluctuations in the new phase are then local phenomena with a short lifetime [21]. With system size decreasing, the energy difference between the stable and metastable phases decrease, so there is a transition range $\Delta T_{tr}$ of temperatures where the probability to find the equilibrium system in any of phases is substantial. However, the equilibration time to achieve this two-phase distribution is an exponential function of the inter-phase barrier energy, and may appear too large for observation. The energy barriers keeping the system in a phase decrease with decreasing particle number $N$, and in small enough systems, a reverse transition from the new to the old phase can be observed and equilibration achieved on experimental time-scales. Then, for these systems heterophasic fluctuations become an observable form of equilibrium thermal motion. The size of the system to observe these fluctuations depends on available observation time; for typical experimental conditions, the characteristic size is in nanometer length-scale. Below, we present a simple model for these fluctuations, and discuss conditions for their observation.

Consider a finite system in thermodynamic equilibrium under conditions of constant particle number $N$, pressure $P$, and temperature $T$ ((NTP) system). Near a phase transition, two-phase states with $N_{new}$ particles in the new phase and $N_{old} = N - N_{new}$ particles in the old phase appear as fluctuations. The fraction of particles belonging to the interface separating phases is neglected. The probability $p(N_{new})$ of finding a state with $N_{new}$ particles in the new phase is defined by the minimal work $W_{min}$ of creating this non-equilibrium state[23]; for an (NTP)-system, this minimal work is given by the non-equilibrium Gibbs free energy $G(N_{new})$ [23,24]. We consider a two-phase model for a finite system with $G(N_{new})$ written as a sum of contributions of one-phase regions and the interface[1,2]:

$$p(N_{new}) \sim e^{-\frac{W_{\min}(N_{new})}{k_B T}},$$
$$W_{\min}(N_{new}) = G(N_{new}) = N_{new}\mu_{new} + (N - N_{new})\mu_{old} + \alpha \Sigma(N_{new}). \quad 1$$

In this formula, $k_B$ is the Boltzmann constant, $\mu_{new}$ and $\mu_{old}$ are the chemical potentials of the new and old phases, $\alpha$ the surface tension, and $\Sigma$ the area of the interface. One assumes that all characteristics except for $N_{new}$, including the shapes of regions occupied by phases, are defined by the condition of minimal Gibbs energy at given $N_{new}$. Note that when the phase-separating interface has nonzero curvature, pressures in phases differ due to the Laplace pressure.

In the macroscopic limit $N \rightarrow \infty$, the condition $\mu_{new} = \mu_{old}$ defines the phase transition line $T = T_\lambda(P)$ in the (T-P)-thermodynamic plane. For finite systems, this condition defines the conventional transition temperature. Near the phase transition temperature along an isobar, the difference $(\mu_{new} - \mu_{old}) \approx -s\Delta T$, where $s = -\partial(\mu_{new} - \mu_{old})/\partial T$ is the per-particle transition entropy, and $\Delta T = T - T_\lambda(P)$. Chemical potentials and surface tension depend on system size. This results in a size-dependent shift of the transition range. One expects these effects to be small when the size of the system is much larger than the interaction radius of particles. In anisotropic systems, surface tension is a local characteristic depending on the local orientation of the interface. Generalization of our model to include anisotropy is rather straightforward, but it involves new parameters. Qualitatively, effects discussed in the current paper are determined by the features of the Gibbs energy landscape on the $N_{new}$ axis: deep minima at the ends, and a maximum between minimum points. These features are rather general for finite systems near phase transitions. These features are schematically illustrated by Fig.1.

The Gibbs energy (1) has two endpoint minima: $G=G_{new}$ at $N_{new}= N$ and $G=G_{old}$ at $N_{new}=0$. In the Gibbs energy landscape on the $N_{new}$ – axis, each of these minima is the bottom of an energy basin associated with corresponding phase. At the endpoints, the interface area $\Sigma(N_{new})$ and the interface contribution to Gibbs energy vanish. The difference in Gibbs energies between the minima is $G_{new}-G_{old} = N(\mu_{new}-\mu_{old})\approx - Ns\Delta T$. Between the endpoints, there is a maximum $G_{max}$ of the Gibbs energy at $N_{new,max} = xN$ ($1\geq x\geq 0$). The maximum separates the basins, and can be found from the equations

$$\frac{dG(N_{new})}{dN_{new}} = \alpha \frac{d\Sigma}{dN_{new}} - (\mu_{old} - \mu_{new}) = 0;$$
$$\frac{d\Sigma}{dN_{new}} = \frac{\mu_{old} - \mu_{new}}{\alpha} \sim \frac{s\Delta T}{\alpha}$$

In a macroscopic system, these equations describe a critical nucleus[23] of the new phase; the radius of this nucleus is $R_c\sim[\alpha/(s\Delta T)]$, the number of particles $N_{new,c} \sim[\alpha/(s\Delta T)]^D$, and the interface area $\Sigma(N_{new})\sim (N_{new})^{(D-1)/D}$, where D=2 or 3 is the number of space dimensions. At $\Delta T=0$, both the critical radius and the related surface energy are macroscopically large, so there is an infinitely high excitation energy barrier between phases. In a finite system, there are finite maximum values for the size, particle number, and interface area of a nucleus, and thus the energy barrier between phases is always finite. For periodic boundary conditions, the highest maximum at $N_{new}=N/2$ is reached at $\Delta T=0$. Note that by definition of the new and old phases $\mu_{old} \geq \mu_{new}$, so the second line in (2) implies $d\Sigma(N_{new})/dN_{new} \geq 0$ and thus $N_{new,c} \leq N/2$.

The transition range of temperatures $\Delta T_{tr} \sim k_B T_\lambda/(sN)$ is defined by the condition that the difference in Gibbs energies between phases is small enough for the equilibrium ensemble to include significant fractions of both phases. However, this definition assumes complete equilibrium. In a large enough system, the equilibration time for this ensemble can exceed available times. This time is determined by the energy of creating the critical nucleus. In the critical range of temperatures, the critical nucleus occupies a significant part of the system, and its size is about the system size, so both the surface area of the interface and the energy to create this nucleus grow with increased system size. A supercooling or overheating beyond the transition range results in a smaller critical nucleus, and accelerates the transition to the new phase, but then the barrier for the transition back to the old phase includes the energy difference between phases, and becomes too large for this transition to happen at experimental time-scales. By diminishing the number $N$ of particles in the system, one can arrive at a system size when for the transition range of temperatures the transition from old to new phase happens on observable times. Then, the difference between excitation energies for a transition from new to old phase and for the transition from old to new phase become small, and both transitions become observable.

In the macroscopic limit, there is no transition range of temperatures; on approaching the phase transition temperature, activation energies for an inter-basin fluctuation $G_{max} - G_{new}$ and $G_{max} - G_{old}$ become macroscopically large, so the probability to find the system close to the basin-separating maximum becomes negligibly small. Each phase can be then defined as an ensemble of states belonging to the corresponding basin. For a finite system in the transition range of temperatures, this phase-basin correspondence becomes conventional: one identifies the phase of the fluctuating system with a basin by convention that for $N_{new} < N_{new,max}$ the system is in the old phase, otherwise the system is in the new phase. The equilibrium ensemble gives a finite probability to find the system in each of the phases. This two-phase interpretation, based on the phase-basin correspondence, becomes ambiguous when there is a significant probability to find the system close to the basin-dividing barrier.

The ambiguity is negligibly small when $G_{max} - G_{new}$ and $G_{max} - G_{new}$ are large compared to thermal energy $k_B T$. The same energies define the average times $\theta_{old}$ and $\theta_{new}$ for the system to continuously keep in same basin; according to the theory of thermally activated processes, these times can be described by an Arrhenius-like formula:

$$\theta_{old} = \frac{1}{\nu} e^{\frac{G_{max} - G_{old}}{k_B T}}, \quad \theta_{new} = \frac{1}{\nu} e^{\frac{G_{max} - G_{new}}{k_B T}},$$
$$G_{max} - G_{old} = \alpha \Sigma_{max} - (N - N_{new})\mu_{old},$$
$$G_{max} - G_{new} = G_{max} - G_{old} - s N \Delta T$$

         **3**

with $w = 1/\tau$ being the frequency of attempts to change the basin; one expects that the time $\tau$ is of the order of the vibration period for system particles.

In a metastable macroscopic system, $\theta_{old}$ is finite and size-independent while $\theta_{new}$ is too large to be observed. The activation energies and the lifetimes $\theta_{new}$ and $\theta_{old}$ decrease with decreasing particle number $N$. The condition that both lifetimes are of the same order of magnitude coincides with the condition defining the transition range $\Delta T_{tr}$ that there is a significant probability of finding the equilibrium system in any of the phases. Then, in a small enough system in the transition range $\Delta T_{tr}$ of temperatures, both the metastable-stable and the stable-metastable transitions can occur at observable times, and random phase changes (heterophasic fluctuations) become a form of thermal motion. The sum $\theta_{new} + \theta_{old}$ gives the average period for these fluctuations. This period depends on $N$ and $\Delta T$; for a given particle number N, this period as a function of temperature has a minimum when $\Delta T=0$ and $\theta_{new}=\theta_{old}$. The minimal period rapidly increases with increasing particle number. The condition $\theta > \theta_{new} + \theta_{old}$ sets the upper limit $N(\theta)$ for particle number in the system to observe heterophasic fluctuations at time-scale $\theta$.

With diminishing system size and all other factors kept constant, $N_{new,max}$ increases towards its largest possible value $N_{new}/2$. At $\Delta T=0$, $G_{max} - G_{old} = G_{max} - G_{new} = \alpha \Sigma_{max} \sim (N)^{(D-1)/D}$. Using (3) one gets the upper limit $N_{up}(\theta)$ for the particle number in a system fluctuating at $\Delta T=0$ between the old and new basins at time-scales $t<\theta$:

$$N_{up}(\theta) \sim \left(\frac{1}{p} \ln \frac{\theta}{\tau}\right)^{D/(D-1)}, \quad p = \frac{r_0^2 \alpha}{k_B T},$$
$$G_{max} - G_{old} = G_{max} - G_{new} = \alpha \Sigma_{max} \sim r_0^2 \alpha N^{\frac{D-1}{D}} \sim k_B T_\lambda \ln \frac{\theta}{\tau} \quad 4$$

$r_0$ is the inter-particle distance. The ratio $p=r_0^2\alpha/(k_B T_\lambda)$ is a non-dimensional materials characteristic.

When $\Delta T \neq 0$, the old phase is metastable. The average lifetime $\theta_{new}$ of the system in the new phase is now larger than that in the old phase. The ratio $\theta_{new}/\theta_{old}=\exp[(G_{old}-G_{new})/$

$k_BT] \geq 1$ is of the order of unity in the transition range of temperatures, and becomes very large beyond this range. Increasing supercooling or overheating of the old phase results in increasingly asymmetric heterophasic switches: the system stays a long time in the new phase and for a short time visits the old one. When $\theta_{new}$ becomes larger than the observation time, the transition is considered irreversible. For a system yielding the condition (4), the reverse transition is observable at time-scale $\theta$ in the temperature range $\Delta T_{hp}(\theta)$

$$\Delta T_{hp}(\theta) = \frac{k_B T_\lambda}{sN} \ln\frac{\theta}{\tau}. \qquad 5$$

Note that the transition range of temperatures $\Delta T_{tr}$ is defined by the condition that in the equilibrium ensemble, the probabilities to find the system in the old as well as in the new phase are of the order of unity; this equilibrium can be reached only at times much larger than $\theta=\theta_{new}+\theta_{old}$. The ratio $\theta/\tau$ may be very large, so the range $\Delta T_{tr}(\theta)$ of temperatures where the heterophasic fluctuations are observable may be larger than the transition range $\Delta T_{tr} \sim k_B T_\lambda/(sN)$, but the asymmetry of phase change rapidly increases when $|\Delta T|>\Delta T_{tr}$

With decreasing particle number $N$, the average period for heterophasic fluctuations decreases, but the probability $P_{inter} \sim \exp[-\alpha\Sigma_{max}/k_BT] \sim \theta/\tau$ of finding the system close to the activation barrier increases. In a small enough system, the phase-basin correspondence and the description of the system in terms of phases can become ambiguous. The condition $P_{inter} \sim \theta/\tau >> 1$, assumed by the two-phase approach, limits the system size from below; when this condition is violated, the two-phase picture is not applicable.

An (NTP)-system large enough to justify the phase-basin correspondence is with a high probability occupied by the old or the new phase. In contrast to that, a macroscopic or very large (NTV)-system (constant $N$, $T$ and volume $V$) has a range of densities $n=N/V$ where two coexisting phases coexist at equilibrium while having different densities. The number $N_{new}$ of particles in the new phase, and thus volumes occupied by phases are determined by the condition that the pressure in the system equals the phase transition pressure $P=P_\lambda(T)$. The minimal work of preparing a state with non-equilibrium value of $N_{new}$ is now the free energy $F(N_{new})=G(N_{new})-PV$, it has a deep minimum (see a schematic plot of $F(N,T,V,N_{new})$ in Fig.2) at the equilibrium value of $N_{new}$. There are also two endpoint minima, both metastable, describing nuclei of corresponding phases. When the particle number $N$ is decreased, the free energy landscape becomes more shallow, and fluctuations of the relative numbers x= $N_{new}/N$ and 1-x= $N_{old}/N$ increase. The phase coexistence picture is justified by the same condition $\theta/\tau>>1$ as the phase-basin approximation.

The simple model of a finite system under periodic boundary condition, used above to discuss system size limitations for observation of heterophasic fluctuations, does not account for many factors. A more realistic model has to include additional system-specific fluctuating characteristics, for example local anisotropy in crystals and liquid crystals. However, one expects that under periodic boundary conditions, the two-phase picture and main qualitative predictions of the simple model still apply.

Boundary conditions other than periodic introduce a new component, the external boundary of the system, which needs to be included in the model. The external boundary contribution to the Gibbs free energy describes a thin layer near the surface where the local structure is perturbed by the presence of the boundary and possible direct interactions with the imbedding matrix. The width $\delta r$ of this layer is expected to be about the width of the phase-separating interface, and assumed to be much smaller than the system size, so the bulk contribution is much larger than the surface contribution. However, close to a phase transition the difference in Gibbs energy between phases is also small, so the surface contribution may become important.

In a system with a boundary, a nucleus can have part of its surface at the interface between phases, and part at the external boundary. The surface tension at the boundary depends on the phase of the adjacent material, so for a two-phase state there are the surface tensions $\alpha_{new}$ and $\alpha_{old}$ for the new and old phases at the external boundary and the surface tension $\alpha$ at the interface between phases. The minimal work to create a critical nucleus at the boundary is usually smaller and thus the probability of nucleation at the surface (heterogeneous nucleation) is much larger than that in the bulk. The difference $w=\alpha_{old}-\alpha_{new}$ characterizes the surface-related bias towards the new phase.

Consider a two-phase state with $N_{new}$ particles in the new phase, $0<N_{new}<N$, and external boundary parts belonging to the new and old phase meeting at the junction with the phase-separating interface. The surface part of the Gibbs energy includes now the energies of the interface and all parts of the external surface and of their junction. The geometry of the two-phase system at given $N_{new}$ is determined by the condition of minimum Gibbs energy. In particular, the angles of contact between surfaces at their junction are determined by the condition of mechanical equilibrium [23]. This, in turn, determines the shape of the two-phase system. The Gibbs energy can be then calculated from the system geometry defined by the angles of contact and $N_{new}$. For the general case, the formula for the Gibbs free energy becomes rather cumbersome, although the dependence of this energy on $N_{new}$ is qualitatively similar to that of a system with periodic boundary conditions. The main feature of this dependence is that in the transition range of temperatures there are two endpoint minima corresponding to the stable and metastable phase, and a maximum describing the barrier energy $G_{max}$ for heterophasic fluctuations.

With bias $w=\alpha_{old}-\alpha_{new} >0$ towards the new phase increasing, the new phase region becomes more and more layer-like. The conventional phase transition temperature, and the phase transition range of temperatures shift towards the old phase due to surface energy gain in transition; this effect is proportional to the fraction of the system at the surface and thus is larger for smaller systems. The energy barrier for nucleation decreases and the frequency of heterophasic fluctuations at transition range temperatures increases. For a large bias $w>\alpha$, no contact angles can satisfy the mechanical equilibrium condition at the junction; in this special case, the entire external surface is always in the new phase, because this lowers the Gibbs energy by at least the energy $\delta G=(W-\alpha)\Sigma(0)$ of creating a double-layer with the new phase at the surface. Here, $\Sigma(0)$ is the external surface area. Below, we discuss this case in more details.

For $w>\alpha$, the minimum of the Gibbs free energy at given $N_{new}$ assumes that the external boundary always belongs to the new phase, and the interface area $\Sigma(N_{new})$ is a monotonously decreasing function of $N_{new}$, with a maximum $\Sigma(0)$ at $N_{new}=0$. Consider a spherical 3D-particle of radius $R$, $\Sigma(0)=4\pi R^2$. The old phase occupies a sphere of radius $R_{old}$ so that its surface area and volume are $\Sigma(N_{new})=4\pi R_{old}^2$ and $V_{old}=(4/3)\pi R_{old}^3$. The number of particles in the new phase is $N_{new}=(N-N_{old})=(N-n_{old}V_{old})$, where $n_{old}$ is the particle number density of the old phase. The Gibbs free energy for the spherical system has the form

$$G(N_{new}) = G(N) + 4\pi\alpha[\frac{3}{4\pi n_{old}}(N - N_{new})]^{2/3} - (N - N_{new})s(T_\lambda - T). \qquad 6$$

Here, *G(N)* is the Gibbs free energy for $N_{new} = N$. The conventional (bulk) transition temperature $T_\lambda$ is defined by the condition $\mu_{new} = \mu_{old}$, and the higher temperature phase is the new phase (for the opposite choice, the sign of the last term in (6) has to be changed). Note that due to different densities of phases, the external radius R depends on $N_{new}$. Below, we assume that this density difference is small, and neglect this effect.

The Gibbs energy (6) is schematically shown in Fig.3. The slope of the interface energy is negative and has a singularity at $N_{new}=N$. Deep in the temperature range ($T<T_\lambda$) of the old phase, the bulk part of the Gibbs energy as a function of $N_{new}$ has a large constant positive slope. The Gibbs energy has endpoint minima at $N_{new}=0$ and $N_{new}=N$. and a maximum at $N_{new,max}$,

$$N_{new,max} = N - \frac{4\pi}{3n_{old}^2}[\frac{2\alpha}{s(T_\lambda - T)}]^3 . \qquad 7$$

The two minima of the Gibbs energy coincide when $T = T_{\lambda new}$,

$$T_{\lambda,new} = T_\lambda - \frac{4\pi\alpha}{sN}(\frac{3N}{4\pi n_{old}})^{\frac{2}{3}},$$

$$N_{new}(T_{\lambda,new}) = N[1-(\frac{2}{3})^3] \sim 0.70N \qquad 8$$

The temperature $T_{\lambda new}$ is the new phase transition temperature; in the macroscopic limit $N\to\infty$ the shift $\Delta T_\lambda = T_\lambda - T_{\lambda new}$ of the phase transition temperature towards the biased phase vanishes as $N^{-1/3}$. With temperature increasing towards $T_\lambda$, the maximum position $N_{new,max}$ shifts to smaller values, reaching zero at $T=T_{sp}$,

$$T_{sp} = T_\lambda - \frac{3\alpha}{s}(\frac{4\pi}{3n_{old}^2 N})^{1/3} \qquad \qquad 9$$

At temperatures $T > T_{sp}$ the Gibbs energy becomes a monotonously decreasing function of $N_{new}$ with only one minimum at $N_{new} = 0$ corresponding to the entire system in new phase.

As in the case of a system in periodic boundary conditions, the form of the Gibbs energy with two endpoint minima and a maximum is the condition for random heterophasic fluctuations as a form of thermal motion observable in small enough systems, with the average period of these fluctuations is given by the formula (3). A fluctuation from the old to the new phase moves the interface from the external surface towards center of the system through the critical configuration at maximum of the Gibbs energy; a fluctuation from new to old phase involves nucleation of an old phase nucleus in the new phase and its fluctuational growth to reach the same critical configuration. The critical configuration is in both cases at $N_{new} = N_{new,max}(T)$. At the new transition temperature $T_{new,\lambda}$ the activation energy $\Delta G$ for this critical configuration from any of phases is

$$\Delta G = G_{max} - G(N) = 4\pi\alpha(\frac{3N}{4\pi n_{old}})^{\frac{2}{3}}[(\frac{2}{3})^2 - (\frac{2}{3})^3] \sim 0.72\alpha(\frac{N}{n_{old}})^{\frac{2}{3}}. \qquad 10$$

At the phase transition temperature, excitation energy to reach the critical configuration is, due to the definition of this temperature, the same from both phases. This excitation energy is lower in a system with a boundary, and lowers with increasing bias.

The melting - crystallization transition in finite systems is frequently studied in experiments and computer simulations. Anisotropy of the crystalline packing makes the surface energy depending on the direction of the interface relative to the anisotropy. The shape of the nuclei is then not spherical (not circular for 2D-systems). Generalization of the model to include anisotropy is rather a straightforward task. Heterophasic oscillations are determined by qualitative features of the Gibbs energy that are characteristic for all finite systems in the phase transition range of temperatures.

In ref. [16], density oscillations with periods $\sim 10^3 \tau$ in a two-dimensional (NTP)-system of $N$=4096 Lennard-Jones particles have been reported. Oscillations have not been seen, and the transition became irreversible at the available simulation time-scale when the size of the system was increased to 36864 or more particles. However, the mechanism of the observed oscillations and particularly the nature of transient states need clarification. The small difference between solid and liquid states in finite 2D-systems can lead to a very wide interface, resulting in a different type of transient states in a small system.

The small width of the interface compared to the system size is a necessary condition for the above two-phase models. For phase transitions characterized by large discontinuity of properties, the interface is known to be only few interparticle distances wide, so the two-phase picture is expected to hold at least qualitatively when N>>1. However, in some systems, for example in systems that are close to a critical point, the differences between phases are small and phase transitions are almost continuous. Data presented in [17] and other publications suggest that in the 2D Lennard-Jones system the width of the crystal-melt interface is much larger than the inter-particle distance. Then, a system with particle number $N$>>1 (for example $N$=2500) may still be too small to host a critical nucleus. In this case, a spatially homogeneous phase change involving excited local structures may have lower excitation energy than the two-phase mechanism.

We thank the NSF-MRSEC program for support of this work. Useful remarks by Neil Snider, and helpful discussions with Antoni Mitus and Rafal Orlik (Technical University Wroclaw, Poland) of computer simulations and melting in 2D Lennard-Jones systems are greatly appreciated.

**Figure captions**

Fig.1. (NTP)-system: a schematic plot of the non-equilibrium Gibbs energy $G(N_{new})$ (solid line), and of the interface and bulk contributions (dashed lines).

Fig.2. (NTV)-system: a schematic plot of the non-equilibrium free energy $F(N_{new})$ (solid line), and of the interface and bulk contributions (dashed lines).

Fig.3. Surface-biased (NTP)-system: non-equilibrium Gibbs energy $G(N_{new})$ (solid line); interface and bulk contributions (dashed lines).

**Figures**

Fig.1

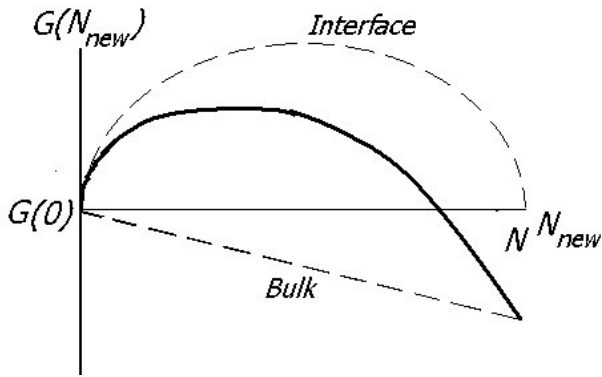

Fig2.

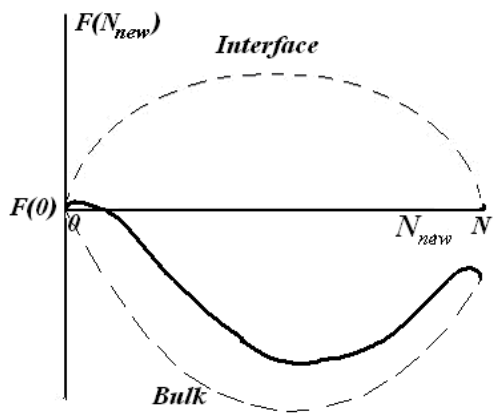

Fig. 3

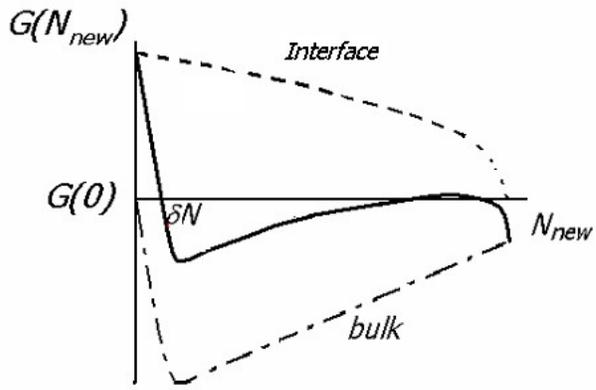